\documentclass[apsrev4-1,prb,superscriptaddress,longbibliography,twocolumn]{revtex4-1}
\usepackage{colortbl}
\usepackage{color}
\usepackage{amsfonts,amsmath,amssymb}
\usepackage[dvipsnames]{xcolor}
\usepackage{tabularx,hhline,tikz,multirow,array}
\usepackage{bm,enumitem}
\usepackage{dcolumn,stmaryrd}
\usepackage[colorlinks=true,pdfborder={0 0 0},linkcolor=red,citecolor = blue]{hyperref}

\newcommand{\Sp}{\mathbf{S}}

\def\rv {{\bf r}}

\def\qv {{\bf q}}

\def \bea {\begin{eqnarray}}
\def \eea {\end{eqnarray}}

\allowdisplaybreaks
\begin{document}
\title{Spontaneous antiferromagnetic skyrmion/antiskyrmion lattice and spiral spin liquid states in the frustrated triangular lattice}

\author{M. Mohylna}
\affiliation{Department of Theoretical Physics and Astrophysics, Institute of Physics, Faculty of Science, Pavol Jozef \v{S}af\'arik University in Ko\v{s}ice, Park Angelinum 9, 041 54 Ko\v{s}ice, Slovak Republic}

\author{F. A. G\'omez Albarrac\'in}
\affiliation{Instituto de F\'isica de L\'iquidos y Sistemas Biol\'ogicos (IFLYSIB), UNLP-CONICET, Facultad de Ciencias Exactas, La Plata, Argentina}
\affiliation{Departamento de F\'isica, Facultad de Ciencias Exactas, Universidad Nacional de La Plata, La Plata, Argentina}
\affiliation{Departamento de Ciencias B\'asicas, Facultad de Ingenier\'ia, Universidad Nacional de La Plata, La Plata, Argentina}

\author{M. \v{Z}ukovi\v{c}}
\affiliation{Department of Theoretical Physics and Astrophysics, Institute of Physics, Faculty of Science, Pavol Jozef \v{S}af\'arik University in Ko\v{s}ice, Park Angelinum 9, 041 54 Ko\v{s}ice, Slovak Republic}

\author{H. D. Rosales}
\email{rosales@fisica.unlp.edu.ar}
\affiliation{Instituto de F\'isica de L\'iquidos y Sistemas Biol\'ogicos (IFLYSIB), UNLP-CONICET, Facultad de Ciencias Exactas, La Plata, Argentina}
\affiliation{Departamento de F\'isica, Facultad de Ciencias Exactas, Universidad Nacional de La Plata, La Plata, Argentina}
\affiliation{Departamento de Ciencias B\'asicas, Facultad de Ingenier\'ia, Universidad Nacional de La Plata, La Plata, Argentina}

\date{\today}

\begin{abstract}
Magnetic skyrmions are topological quasiparticles of great interest for data storage applications
because of their small size, high stability, and ease of manipulation via electric current. 
Antiferromagnetic (AF) skyrmions, with new features and huge benefits (ultra-small skyrmion sizes, no transverse deflection and efficient manipulation), have recently become the subject of intense focus. Here we show that a spontaneous antiferromagnetic skyrmion/antiskyrmion lattice (AF-SkL/ASkL) emerges in the classical Heisenberg antiferromagnet  on the triangular-lattice under magnetic fields, taking only exchange interactions up to third nearest neighbors ($J_1$-$J_2$-$J_3$). By means of the Luttinger-Tisza approximation and  large-scale Monte-Carlo  simulations (combining Parallel-Tempering and overrelaxation with the Metropolis algorithm), we present a  rich $J_2$-$J_3$ magnetic phase diagram including exotic multiple-q phases, degenerate states and a spontaneous AF-SkL/ASkL lattice at intermediate magnetic fields. In addition, we show that at zero magnetic field, exotic spin liquid states with ring-like degeneracy emerge at intermediate temperatures, which are broken by thermal fluctuations selecting different multiple-q states. These findings greatly enrich the research on antiferromagnetic skyrmions in centrosymmetric materials or lattices including relatively weak Dzyaloshinskii-Moriya interaction.
\end{abstract}

\maketitle

\section{Introduction}

Since the experimental discovery of magnetic skyrmion crystals in MnSi\cite{MnSi}, magnetic skyrmions -  noncoplanar spin configurations with  nonzero topological  number- have attracted great research interest due to their high stability and particle-like behavior. Their particular topological properties, small size and  unique dynamic behavior render these magnetic textures  promising candidates for potential applications to next-generation spintronics devices \cite{NagaosaTech,SampaioTech,FertTech}.  In this context, antiferromagnetic skyrmions \cite{Ezawa2016} have become the focus of intense work, since in these textures the "skyrmion Hall effect" \cite{SkyHall1,SkyHall2} should be suppressed \cite{SkyAFHall1,SkyAFHall2}. This has been further supported by experimental evidence in ferrimagnetic skyrmions \cite{SkyFerri} and antiferromagnetic bubbles \cite{SkyAFBubbles}. Antiferromagnetic skyrmion-like textures have been realized in synthetic antiferromagnets \cite{AFSky1}, and materials $\alpha-$Fe$_2$O$_3$ \cite{AFSky2} and MnSc$_2$S$_4$ \cite{Gao2020,Rosales2022}. 

In a large number of cases, periodic arrays of magnetic skyrmions are stabilized when an external magnetic field is applied in non-centrosymmetric systems displaying the antisymmetric  Dzyaloshinskii-Moriya interaction (DMI) \cite{DM1,DM2}.  In these   systems, a  ferromagnetic exchange interaction competes with DMI, which induces stability of periodic arrangement of helical spin structures.  In addition to this, recent studies have revealed that not only skyrmions but other topological spin textures \cite{BeyondSky}, such as antiferromagnetic skyrmion lattices \cite{Rosales2015,Osorio1,Osorio2,Zukovic1,Zukovic2,villalba,Zukovic3,Diep2022}, can be  stabilized even in centrosymmetric lattices through different mechanisms as  exchange frustration \cite{Okubo2012}, bond-dependent exchange anisotropy \cite{Gao2020,Amoroso2020,Wang2021,Hayami2021-a,Utesov2021,Yambe2021,Amoroso2021,Hayami2022-a}, the Ruderman-Kittel-Kasuya-Yosida (RKKY) interaction in itinerant magnets\cite{WangRKKY2020},  higher-order exchange interactions\cite{Paul2020}, etc. It has been shown that in these types of antiferromagnetic skyrmion lattices, formed for example by three interpenetrated triangular sublattices, an external magnetic field may tune the topological Hall effect \cite{Tome2022}. The purpose of this investigation is to explore the first mechanism, where skyrmion crystals may emerge by incorporating the effect of thermal fluctuations in frustrated systems. In fact, the presence of the skyrmion phase in centrosymmetric frustrated magnets in the absence of the DMI was confirmed experimentally very recently\cite{Kurumaji2019}. On the theoretical side, an example of this phenomena was studied by Okubo et al\cite{Okubo2012} in which the authors show that at finite magnetic field and temperature in the $J_1$-$J_2$ (or $J_1$-$J_3$) Heisenberg model on the triangular-lattice, a specific configuration  of magnetic frustration  (ferromagnetic $J_1$, antiferromagnetic  $J_{(2,3)}$ with $J_{(2,3)}>>|J_1|$) induces a spontaneous ferromagentic skyrmion/antiskyrmion crystal. Here, antiskyrmions are magnetic structures analog to skyrmions, but with opposite topological charge, and could thus be considered their ``antiparticles''\cite{leonov2017edge}. Antiskyrmions have been realized in materials such as  Schreibersite (Fe,Ni)$_3$P \cite{antiskyMat1},  in Fe/Gd-based multilayers \cite{heigl2021dipolar}, and have been known to be stabilized in models including spin-orbit coupling \cite{Kumar1,Kumar2} and  layer-dependent DMI \cite{Hayami1}. 

In addition to the skyrmion lattice formation in a frustrated system \cite{Leonov2015}, another fundamental concept that emerges in these systems is that of  spiral spin liquids (SSL). Here, the ground state configurations form a continuous manifold in reciprocal space, that strongly governs the low-temperature physics. 
Some examples of systems showing SSL are the Heisenberg model on the square\cite{Chandra1988}, honeycomb\cite{Mulder2010,yao2021generic} lattices, and an approximate version of this phase has been experimentally identified in the spin-$5/2$ diamond lattice compound MnSc$_2$S$_4$\cite{gao2016} and more recently in the van der Waals honeycomb magnet FeCl$_3$\cite{gao2022spiral}. In general, it is well know that degeneracy enhances quantum fluctuations\cite{balents2010spin}; therefore, classical spin liquids are excellent candidates to realize the quantum version at low temperatures.

In the present study, we show that magnetic frustration in the pure classical antiferromagnetic $J_1$-$J_2$-$J_3$ Heisenberg model on the triangular lattice, induces the emergence of a spontaneous antiferromagnetic AF-SkL/ASkL at moderate values of $J_2/J_1$, $J_3/J_1$ and external magnetic field. Furthermore, at zero magnetic field two kinds of exotic spin-liquid states\cite{Shimokawa2019}  emerge according to the  $J_3/J_2$ ratio.
To this end, we employ two complementary approaches: the Luttinger-Tisza approximation (LTA)\cite{Luttinger1946,Luttinger1951} to explore the $T=0$ $J_2$-$J_3$  ground state phase diagram, and large scale classical Monte-Carlo (MC) simulations of the spin Hamiltonian  (combining Parallel Tempering and overrelaxation with the Metropolis algorithm) to incorporate thermal fluctuations and the effect of an external magnetic field. 

The rest of the manuscript is organized as follows. In Sec.~II, we introduce the frustrated spin model and we present the $T=0$  magnetic phase diagram at zero field. Then, we discuss the multiple-q and disordered states. In Sec.~III, we show our simulations analysis at finite temperature, identifying the spin liquid and spontaneous AF-SkL/ASkL regions, which are the central points of our results at zero and finite magnetic field, respectively. In the first case, we focus on the   $J_2 =2J_3$ line, where the model exhibits two types of  degenerate states (spiral spin liquids). In the second case, we study the region  where spontaneous AF-SkL/ASkL phases are stabilized in a  ``pocket'' at finite  temperature and magnetic field. Section IV is devoted to the summary and conclusions.

\section{Model and Zero Temperature Phase Diagram}
We focus on the extended $J_1$-$J_2$-$J_3$ classical antiferromagnetic Heisenberg model on the triangular lattice under a magnetic field, as the simplest model Hamiltonian which incorporates different levels of frustration in this lattice geometry. The Hamiltonian is given by

\begin{eqnarray} \label{eq:H}
\mathcal{H}&=&J_1\sum_{\langle i,j \rangle} \Sp_i\cdot\Sp_j + J_2\sum_{\langle\langle i,j \rangle\rangle} \Sp_i\cdot\Sp_j + J_3\sum_{\langle\langle\langle i,j \rangle\rangle\rangle} \Sp_i\cdot\Sp_j \nonumber\\
&&- B\sum_i\,S_i^z
\end{eqnarray}
\noindent where $\Sp_i$ are unit-vector classical Heisenberg spins, $J_1,J_2,J_3$ are the first-,second- and third-nearest neighbor exchange interactions, $\langle i,j \rangle,\langle\langle i,j \rangle\rangle,\langle\langle\langle i,j \rangle\rangle\rangle$ indicate the sum over first, second and third nearest neighbor pairs, and $B$ is the magnitude of the external magnetic field in the $z$ direction. In this work, we are interested in exploring the effect of competing frustrating interactions, so we take all exchange interactions antiferromagnetic (i.e. $J_{1,2,3}>0$). For simplicity,  we fix  $J_1=1$ as the scale through the rest of this work.

In the triangular lattice, for the nearest neighbor antiferromagnet ($J_2=J_3=0$), at zero external field ($B=0$) the magnetic moments form a 120$^{\circ}$ spin-structure with an  ordering wavevector $\qv^*=(4\pi/3, 0)$ and a trivial six-fold degeneracy related to permutations of the spin triad. At  finite temperature and magnetic field, thermal fluctuations lift the degeneracy of the ground state and stabilize collinear and coplanar states \cite{Gvozdikova2011,seabra2011phase}. 

\begin{figure*}[htb]
\includegraphics[width=1\textwidth]{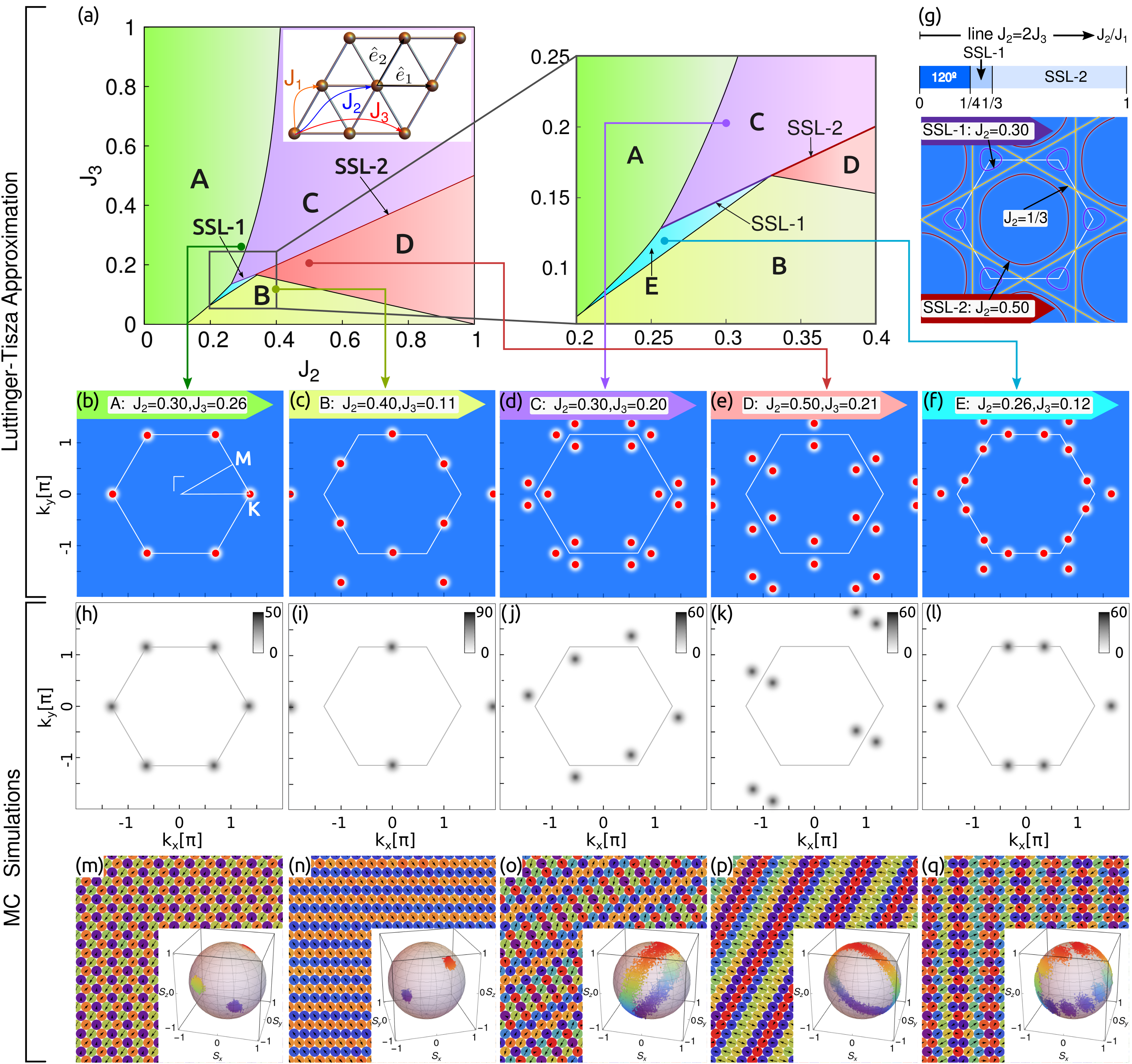}
\caption{\label{fig:PD-LTA} {(Color online) 
LTA (top, panels (a) - (g)) and MC simulations (bottom, panels (h) - (q)) results. LTA: (a) $J_2$-$J_3$ phase diagram obtained from LTA, with five regions {\bf A}-{\bf E} (separated by dotted lines). The inset shows the triangular lattice, lattice vectors $\hat{e}_1,\hat{e}_2$ and the $J_1,J_2,J_3$ couplings. The zoomed area shows phase {\bf E} and its boundaries. SSL-1 and SSL-2, along the line $J_2=2J_3$, correspond to a line with degenerate ground states. Representative ordering wave vectors distributions $\qv^*$ (red dots and colored lines) obtained from LTA are shown in panels (b) to (g), where the values of the couplings are indicated on top and the first Brillouin zone is drawn. In panel (a) we indicate the high symmetry points $\Gamma$, $K$ and $M$.  Each panel corresponds to a different phase, panels (b) to (f) for phases {\bf A} to {\bf E}, and panel (g) presents the two non-equivalent manifolds of classically degenerate spiral states SSL-1 (purple line) and SSL-2 (red line).  
MC simulations: Representative structure factors $\sqrt{S_{\bf q}}$ (panels h-l) and their corresponding real space configurations (panels m-q) for five low temperature ($T=10^{-3}$) different phases, {\bf A}-{\bf E}, where the values of the couplings match those chosen in the LTA. The insets show the spherical snapshots.}}
\end{figure*}

The possible ground states of the  model presented in Eq.~(\ref{eq:H}) at zero magnetic field have been discussed qualitatively in a earlier work by Messio et al. by means of a variational approach \cite{Lhuillier2011}.  Through this technique, three regions in the $J_2-J_3$ phase diagram were identified: a region with coplanar magnetic order (which matches region {\bf A} from the phase diagram presented in Fig.~\ref{fig:PD-LTA}(a)), a region with tetrahedral order ({\bf B}) and a broad region where the results suggested a possible spiral order ({\bf C}, {\bf D} and {\bf E}). In  region {\bf A}, the magnetic structure in each $\sqrt{3}\times\sqrt{3}$ triangular sublattice is ferromagnetic, and the spins from each sublattice are coplanar. The corresponding structure factor is characterized by six peaks in the $K$ points from the Brillouin zone (BZ). For the tetrahedral order (phase {\bf B}), the magnetic unit cell is formed by four spins in the direction from the center to the vertex of a regular tetrahedron. In this order, there are six peaks in the $M$ points from the BZ.

Here we dive further into the model in Eq.~(\ref{eq:H}), first, analyzing in detail all the phases and possible multiple-q states at $B=0$ with the LTA; then, using MC simulations we explore the emergent phases at $B>0$, including a spontaneous AF-SkL/ASkL at finite temperature.

\subsection{Luttinger-Tisza Approximation ($T=0, B=0$)}  \label{subsec:LTA}

To explore the classical ground state phase diagram in the absence of magnetic field at zero temperature, we resort to the Luttinger-Tisza approximation (LTA) (also known as the spherical model)\cite{Luttinger1946,Luttinger1951}. Within this scheme, instead of imposing the local ``strong constraint'' $|\Sp_i|=1$, one imposes a global so-called ``weak constraint'' $\sum_i|\Sp_i|^2=N\,S^2$, where $N$ is the number of lattice sites. With this softer constraint, the model Hamiltonian (\ref{eq:H}) can be diagonalized by a simple Fourier transformation as $S^{\alpha}_{\qv}=\frac{1}{\sqrt{N}}\sum_j\,S^{\alpha}_je^{i\,\rv_j\cdot\qv}$, $\alpha = x, y, z$ is the
spin component and $\rv_j$ and $\qv$ denote the position and pseudo-momentum respectively. The Hamiltonian then becomes

\begin{eqnarray}
\mathcal{H}&=&\sum_{\qv}J(\qv)\,\Sp_{\qv}\cdot\Sp_{-\qv}
\end{eqnarray}
where the sum in $\qv$ runs over all wave vectors in the first Brillouin zone,  $J(\qv)=\sum_{a=1}^3J_a\sum_{\delta_a}\cos(\qv\cdot\delta_a)$ defines the Fourier transforms of the exchange interactions. Here $a=1,2,3$ indicate first, second and third neighbours, respectively, and $\{\delta_1\}\equiv\{\hat{e}_1,-\hat{e}_1+\hat{e}_2,-\hat{e}_2\}$, $\{\delta_2\}\equiv\{\hat{e}_1+\hat{e}_2,\hat{e}_1-2\hat{e}_2,-2\hat{e}_1+\hat{e}_2\}$ and $\{\delta_3\}\equiv\{2\hat{e}_1,2\hat{e}_2,2(-\hat{e}_1+\hat{e}_2)\}$), where $\hat{e}_1=\hat{x}$ and $\hat{e}_2=\hat{x}/2+\sqrt{3}\,\hat{y}/2$ are unit vectors depicted in Fig.~\ref{fig:PD-LTA}(a). The ground-state energy is associated with the lowest value of $J(\qv)$ which defines the ordering wave-vector $\qv^*$.  Within this approximation, we find different solutions characterized by the number and the position of $\qv^*$. By comparing the ground state energy of these ordered states, we construct a phase diagram in the $J_2$-$J_3$ plane shown in Fig.~\ref{fig:PD-LTA}(a). As shown in the figure, we find seven different phases with different distributions of the ordering wave vectors $\qv^*$. Among them, it can be seen that two out of seven phases possess a infinite degenerate $\qv^*$ number. We label these two phases by  SSL-1 and SSL-2, while we denote the other five phases by the letters {\bf A}-{\bf E}. We describe these findings below:

\begin{itemize}
\item Phases labeled as {\bf A} and phase {\bf B}  are in the same region as the previous study \cite{Lhuillier2011}, with  ordering wave vectors, shown in Fig.~\ref{fig:PD-LTA}(b)-(c),  in the $K$ and $M$ point of the BZ. Most importantly, LTA sheds light on the remaining region in the $J_2-J_3$ phase space, defining three separate phases, {\bf C}, {\bf D} and {\bf E}.
\item Phases {\bf C} and {\bf D} (Fig.~\ref{fig:PD-LTA}(d)-(e)) present three incommensurate and inequivalent $\bf{q}^*$ orders, satisfying $\sum_{i=1}^3\bf{q}^*_i=0$, and thus signaling potential triple-q phases with non-zero temperature and magnetic field.  The difference between these two phases lies in the position of the minima: in the {\bf C} phase, they lie in the symmetric lines connecting the $K$ and $\Gamma$ points, while in the {\bf D} phase, in the lines connecting the $M$ and $\Gamma$ points.
\item In phase {\bf E} (Fig.~\ref{fig:PD-LTA}(f)) there are twelve ordering  wave vectors at the border of the BZ.
\item Finally, we identify a special line $J_2=2J_3$  where degenerate momentum vectors form  spiral contours (panel (g) in Fig.~\ref{fig:PD-LTA}). Along this line, for $1/4<J_2$, the minimum energy solutions correspond to $\qv^*$ satisfying the relation 
\begin{eqnarray}
J_1-J_2+2\,J_2\gamma_{\qv}&=&0
\end{eqnarray}
with $\gamma_{\qv}=\sum_{\delta_1}\cos(\qv\cdot\delta_1)$. As can be seen from the previous relation, the spiral wavevector is not uniquely fixed. For $1/4<J_2< 1/3$, we observe several spiral contours around  the $K$ points (red lines). Increasing the value of $J_2$ we arrive at the special point $J_2=1/3$ ($J_3=1/6$) where all the contours  merge on a regular contour that touches the BZ boundary at the $M$ points (yellow lines). For $J_2>1/3$ the spiral contour is a single closed loop around the center in the first BZ  (orange lines). This picture is similar to what happens in the $J_1=2J_2$ case in the honeycomb lattice\cite{Okumura2010,Ganesh2010}.
\end{itemize} 

Therefore, we see that the LTA analysis indicates two types of regions where exotic phenomena may arise when considering the effect of thermal fluctuations under a magnetic field. On the one hand, there is the $J_2=2J_3>1/4$ line, where the lowest bands show a semiextensive degeneracy, suggesting possible  spin liquid behavior. Around this region a quantum chiral spin liquid was found for $S=1/2$ spins \cite{gong2019chiral}. On the other hand, there are two broad regions in parameter space where there are six incommensurate  $\qv^*$ peaks, where skyrmion-like phases may be stabilized. We explore these possibilities through high performance simulations in the following section.

\section{FINITE TEMPERATURE BEHAVIOR}

Motivated by the promising results of the LTA, we wish to investigate the possible emergence of the spin liquid and triple-q (skyrmion-like) states at moderate temperatures under magnetic fields, resorting to two complementary Monte-Carlo methods: Parallel Tempering \cite{Swendsen1986} which has proved to be a powerful tool in the study of the systems with a complex energy surface and the Metropolis algorithm combined with the over-relaxation method \cite{Creutz1987}. 

We run Parallel Tempering simulations for lattice sizes $N=L^2$ with $L=21-126$, using $160-300$ replicas (temperatures), depending on the system size and the region of the phase diagram. The temperature set is chosen to follow the geometrical progression as it improves the replica exchange acceptance rates at low temperatures and with a sufficient number of points still provides reasonable resolution at higher ones. Since the problem is amenable to massive parallelization, the simulations are implemented on General Purpose Graphical Processing Units (GPGPU) using CUDA, which allowed to simulate all the replicas at different field values simultaneously. For each replica we use $5-9\times10^6$ MC sweeps for equilibration and half of that amount for calculating mean values. The replica swapping to Metropolis sweep ratio is $1:1$ and occurs after each Metropolis sweep through the whole lattice.

In the second approach, MC simulations were performed using the Metropolis algorithm combined with overrelaxation (microcanonical) updates. We use an annealing scheme to lower the temperature ($T$) at fixed external magnetic field ($B$). Simulations were performed for $L=12-72$ and periodic boundary conditions. $10^5$-$10^6$  MC steps were used for initial relaxation, and  measurements were taken in twice as many MC steps.

To determine the finite-temperature phase diagram, we measure the specific heat $C=\frac{\langle \mathcal{H}^2\rangle-\langle \mathcal{H}\rangle^2}{NT^2}$, magnetization $M=\frac{1}{N}\langle \sum_iS^z_i\rangle$, magnetic susceptibility $dM/dB$, the chiral order parameter, i.e the total scalar chirality, $\chi_Q=\langle \frac{1}{4\pi}\sum_i\chi_i\rangle$ with $\chi_i=\Sp_{i_1}\cdot(\Sp_{i_2}\times\Sp_{i_3})$ where ($i_1,i_2,i_3$) are the indices of the three sites on every elementary triangle at the site $i$. In addition we compute the  perpendicular $S^{\perp}_{\qv}$, the longitudinal $S^{||}_{\qv}$ and the total $S_{\qv}=S^{\perp}_{\qv}+S^{||}_{\qv}$ spin structure factors with expressions given by   

\begin{eqnarray}
S^{\perp}_{\qv}&=&\frac{1}{N}\left\langle\left|\sum_{j}S^{x}_je^{i\qv\cdot\rv_j}\right|^2+\left|\sum_{j}S^{y}_je^{i\qv\cdot\rv_j}\right|^2\right\rangle\\
S^{||}_{\qv}&=&\frac{1}{N}\left\langle\left|\sum_{j}S^{z}_je^{i\qv\cdot\rv_j}\right|^2\right\rangle
\end{eqnarray}

\noindent where the angle bracket $\langle \cdots \rangle$ represents the thermal average.

\subsection{$J_1$-$J_2$-$J_3$ model ($B=0$)}
%
\subsubsection{\textbf{A-E} phases }
Firstly, we explore the low-temperature phases at zero magnetic field ($B=0$), in order to compare the emergent ({\bf A}-{\bf E}) phases with the LTA phase diagram from Fig.~\ref{fig:PD-LTA}. Typical real space textures  and their corresponding structure factors  are presented in Fig.~\ref{fig:PD-LTA}, panels (h-q). It is important to mention that sharp spots observed in the total structure factor $\sqrt{S_{\bf q}}$, at the ordering wave vectors  positions ${\bf q}^*$, are not true Bragg peaks because we are studying an isotropic two-dimensional Heisenberg model. Thus, it is well known that observed sharp spots are  actually quasi-Bragg peaks associated with power-law spin correlations \cite{Okubo2012,Shimokawa2019}.  In addition, we include an inset showing a spherical snapshot, drawing the spins of the configuration from the center of the sphere. The colors indicate the projection along the magnetic field (red is completely aligned, blue anti-aligned). A first observation is that indeed the low-temperature phases may also be classified in five types of phases, as shown in the LTA analysis. Phase A matches the LTA results: a coplanar three-spin arrangement with ferromagnetic sublattice order, characterized by a structure factor with six symmetric peaks in the BZ (compare Fig.~\ref{fig:PD-LTA}(b) and Fig.~\ref{fig:PD-LTA}(h)). We show in Fig.~\ref{fig:PD-LTA}(m)  the typical spin configuration and the spin structure factors showing sharp peaks at the $K$ points in the BZ. A different situation arises for region {\bf B}. Here, a single-q order, characterized by one of the commensurate wavevector on the BZ edge ($M$ point), with antiferromagnetic stripes emerges; we show an example for one realization in Fig.~\ref{fig:PD-LTA}(n) with the corresponding structure factor, panel (i). This arrangement has the same energy as the proposed ``tetrahedral'' texture, but it is clearly more collinear, therefore being favoured by thermal fluctuations in an order-by-disorder selection \cite{ZhitomirskyTrig,AlbaPujol}. As for regions {\bf C} and {\bf D} ({\bf E}), a non-trivial helical-like single-q (double-q) order emerges characterized by an incommensurate wavevector;  typical snapshots are shown in Figs.~\ref{fig:PD-LTA}(o), (p) and (q). The structure factor agrees with the LTA prediction: $\bf{q}$-peaks are in the $K-\Gamma$ line for region {\bf C} (compare Fig.~\ref{fig:PD-LTA} panels (d) and (j)), in the $M-\Gamma$ line for phase {\bf D} (compare Fig.~\ref{fig:PD-LTA} panels (e) and (k)) and at the border of the BZ for phase {\bf E} (compare Fig.~\ref{fig:PD-LTA} panels (f) and (l)).

\begin{figure*}[ht!]
\includegraphics[width=1.0\textwidth]{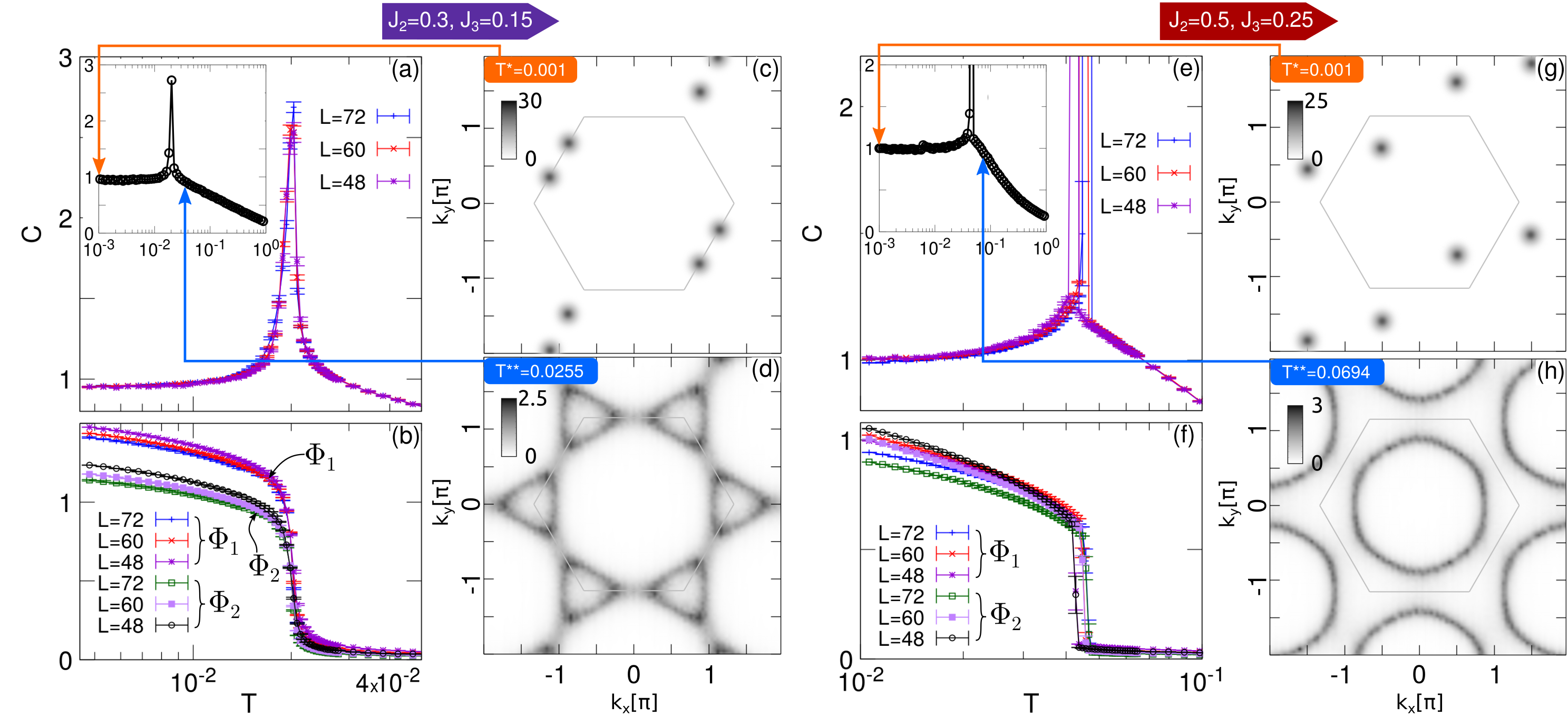}
\caption{\label{fig:SqSL}{(Color online)
Specific heat ($C$, panels (a),(e)) and order parameters $\Phi_{1,2}$ (panels (b),(f)) as a function of temperature for three different system sizes $L=48,60,72$, averaged over $10$ independent MC realizations per size, and  high ($T^{**}$) and low ($T^{*}$) temperature structure factors ($\sqrt{S_{\qv}}$), for two sets of parameters along the $J_2=2J_3$ degenerate line, $J_2=0.3$ (left) and $J_2=0.5$ (right). In the inset of panels (a) and (e), the full range of $C$ vs $T$ is shown for one $L=60$ realization, where  temperatures ($T^*$ and $T^{**}$), corresponding to the depicted structure factors, are indicated with  arrows.}
}
\end{figure*}
\subsubsection{ Degenerate line $J_2=2J_3>1/4$}
We now focus on the degenerate line $J_2=2\,J_3$ in the phase diagram in Fig.~\ref{fig:PD-LTA}. For $J_2\geq 1/4$, following the  LTA analysis, this region may be divided into two types of possibly degenerate ground states, for $J_2<1/3$ and for $J_2>1/3$. In the first case, as shown in Fig.~\ref{fig:PD-LTA}(g), at $T=0$ the energy minima form lines around the $K$ points in the BZ, while in the second case they form a closed loop centered in the $\Gamma$ point of the BZ. To inspect the effect of temperature, we  study several variables for two sets of representative parameters along this line, $J_2=0.3$ and $J_2=0.5$, Fig.~\ref{fig:SqSL}, left, panels (a) - (d), and right, panels (e) - (h), respectively.

First, we inspect the specific heat $C$ as a function of temperature, shown in  panels (a) and (e) for three different system sizes. 
We note that, in both cases, at a given temperature there is a sharp peak in the specific heat and thus a possible phase transition.  Then, we analyze the intensity plots of the spin structure factor $\sqrt{S_{\bf q}}$ just before the peak (at higher $T$) and at the lowest simulated temperature, presented in panels (c),(d),(g) and (h). The selected temperatures are indicated in the $C$ vs $T$ curves in the insets of panels (a) and (e). For $J_2=0.3$ (SSL-1 phase), at high temperature (panel (d)), the system is in a disordered state with spiral and degenerate contour around the $K$ point. Decreasing the temperature further, there is an entropic order-by-disorder (OBD) \cite{OBDspiral} selection of $\qv^*$ sharp peaks at the border of the BZ (panel (c)). A similar entropic phenomenon occurs in the SSL-2 phase ($J_2=0.3$). Here, the system presents a degenerate contour around the $\Gamma$ point at higher temperature (consistent with the LTA results) (panel (h)); while at very low temperatures, after the peak in the specific heat, a couple of sharp peaks at incommensurate $\qv^*$ vectors indicate an OBD selection (panel (g)). The specific heat at low temperatures remains slightly lower than 1 (in units of the Boltzmann constant), which is an indicator of remnant soft modes, that lower the free energy and thus the specific heat per spin is lower than the expected by the equipartition theorem. \cite{KagomeOBD1,KagomeOBD2,Rosales2016}

The OBD selection is associated with the breaking of discrete symmetries, whereas the continuous SO(3) symmetry from the isotropic Hamiltonian remains unbroken, as stated by the Mermin-Wagner theorem. Similar phenomena has been found in the $J_1-J_2$ classical model in the honeycomb lattice \cite{Okumura2010}. To further study this transition, we build two order parameters $\Phi_{a}$ preserving the SO(3) symmetry but describing the $C_3$ lattice-rotational-symmetry breaking in the direction of first ($a=1$) and second ($a=2$) nearest neighbors:

\begin{eqnarray}
\Phi_{a} &=&\frac{1}{N}\left|\sum_{i} \Sp_i\cdot \left(\Sp_{i_1^{(a)}}+\omega\Sp_{i_2^{(a)}}+\omega^2\Sp_{i_3^{(a)}}\right) \right|
\end{eqnarray}

\noindent where $N$ is total number of sites, $\omega=e^{i\frac{2\pi }{3}}$, $i_1^{(a)},i_2^{(a)},i_3^{(a)}$ correspond to the three nonequivalent first ($a=1$) or second ($a=2$) nearest neighbors of spin $\Sp_i$, with relative positions $\delta_{a}$ defined in Sec.~\ref{subsec:LTA}. From the analysis of the structure factors at low temperature (Fig.~\ref{fig:SqSL}, panels (c) and (g)), it can be seen that the directions of the selected $\qv^*$ do not match exactly the first or second nearest neighbor directions, but are a combination of both of them, and thus we expect both $\Phi_{1,2}$ to have non-zero values at low temperature. We then plot both these parameters as a function of temperature in Fig.~\ref{fig:SqSL}(b) and (f), for system sizes $L=48, 60, 72$, averaged over $10$ independent copies. We see that indeed the transition in the specific heat is associated with a jump of $\Phi_{1,2}$ from zero to a finite value, showing that there is a discrete symmetry breaking. We defer the study of the nature of the transitions for future work.

Our findings show that in some geometries, the effect of strong magnetic frustration remains robust, even at high temperatures where thermal fluctuations are assumed to be large. This was observed, for example, in the MnSc$_2$S$_4$ compound \cite{gao2016}. Much more connected with our results, very recently, a possible $U(1)$ SSL state was predicted in the van der Waals magnet FeCl$_3$\cite{gao2022spiral}, where magnetic sites  Fe$^{3+}$ ($S=5/2$) form honeycomb layers (ABC-stacked) along the $c$ axis. In this case, by neutron scattering measurements, the authors found  a continuous ring of scattering around $\Gamma$ providing direct evidence for the existence of a SSL state. This is a quite similar situation to what happens in our SSL-2 phase (see Fig.~\ref{fig:SqSL}, right column).  As a possible experimental realization, we can mention  layers of magnetic transition metals $X$ ($X=$Co, Cr, Fe, Mn), adsorbed onto a monolayer of transition metal dichalcogenides (MoS$_2$, WS$_2$, or WSe$_2$)\cite{Fang2021}.

\begin{figure}[htb]
\includegraphics[width=1.\columnwidth]{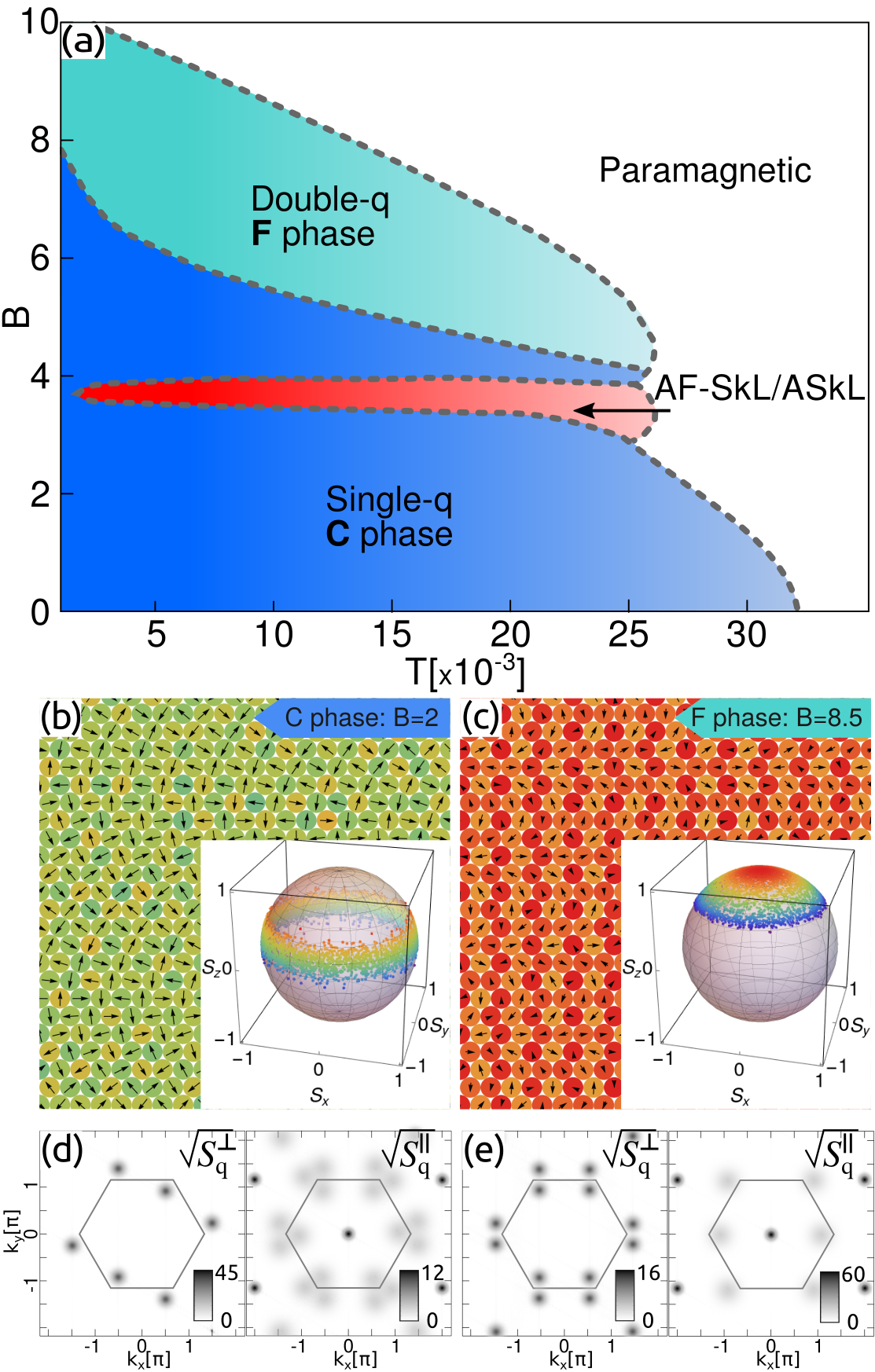}
\caption{\label{fig:PD-asky} {(Color online) (Top - panel (a)) Temperature vs magnetic field phase diagram obtained from simulations for $J_2=0.3$, $J_3=0.16$. A spontaneous antiferromagnetic  skyrmion/antiskyrmion lattice is stabilized in a broad region, indicated in red. At larger fields, a double-q phase ({\bf F}) emerges. (Bottom - panels (b-e)) Typical  spin textures, spherical snapshots and corresponding transverse and longitudinal structure factors $\sqrt{S^{\perp}_{\qv}}, \sqrt{S^{||}_{\qv}}$ for phases stabilized at  $T=10^{-3}$ and different magnetic fields, $B=2$ (panels (b,d), {\bf C} phase), and $B=8.5$ (panels (c,e),  {\bf F} phase).}
}
\end{figure}
%

\subsection{Effect of an external magnetic field: antiferromagnetic skyrmion/antiskyrmion lattice}

In general, topological spin configurations may be classified in terms of their topological charge $Q$ and the helicity $\gamma$ \cite{NagaosaTech}. For example, systems hosting magnetic skyrmions stabilized by the the isotropic DMI, will support ``Neel-type'' or ``Bloch-type'' skyrmions,  depending on the direction of Dzyaloshinsky-Moriya vector which is determined by crystal symmetry. So, the DMI selects skyrmions with $Q=-1$ and fixed helicity as the energetically most favorable spin configuration. However, systems with an isotropic exchange interactions will support any value of both the topological charge  and the helicity. Here we will show that frustrated exchange interactions are capable of stabilizing antiferromagnetic skyrmions or antiskyrmions with topological charge $|Q|=1$ and ``free'' (not fixed) helicity.

In this section we study the possible realization of a frustration-induced SkX state in the model in Eq.~(\ref{eq:H}) at  finite temperature and in the presence of a magnetic field. For this reason, we first inspect the phases that include classical solutions (at $T=0$) with inconmensurable  $\qv^*$ vectors, i.e. phases {\bf C} and {\bf D}.  While in phase {\bf D}, our simulations indicate the stabilization of single-q and double-q phases, the situation becomes completely different in phase {\bf C}, where exotic multiple-q phases are stabilized. We have also checked that the zero-field particular double-q structure found at low temperatures  in  phase {\bf E} does not change significantly with applied field. For these reasons, we focus here on the {\bf C} phase.

As in the previous section, we performed MC simulations for systems with $N=L^2$ ($L=21-126$) sites on a triangular lattice with periodic boundary conditions (Parallel Tempering and standard Metropolis + overrelaxation). To identify the different phases, we calculate the specific heat $C$, the magnetization $M$, magnetic susceptibility $dM/dB$, and the scalar chirality $\chi_Q$, defined in the previous section, combining these parameters to determine the phase boundaries.

\begin{figure*}[htb]
\includegraphics[width=1.0\textwidth]{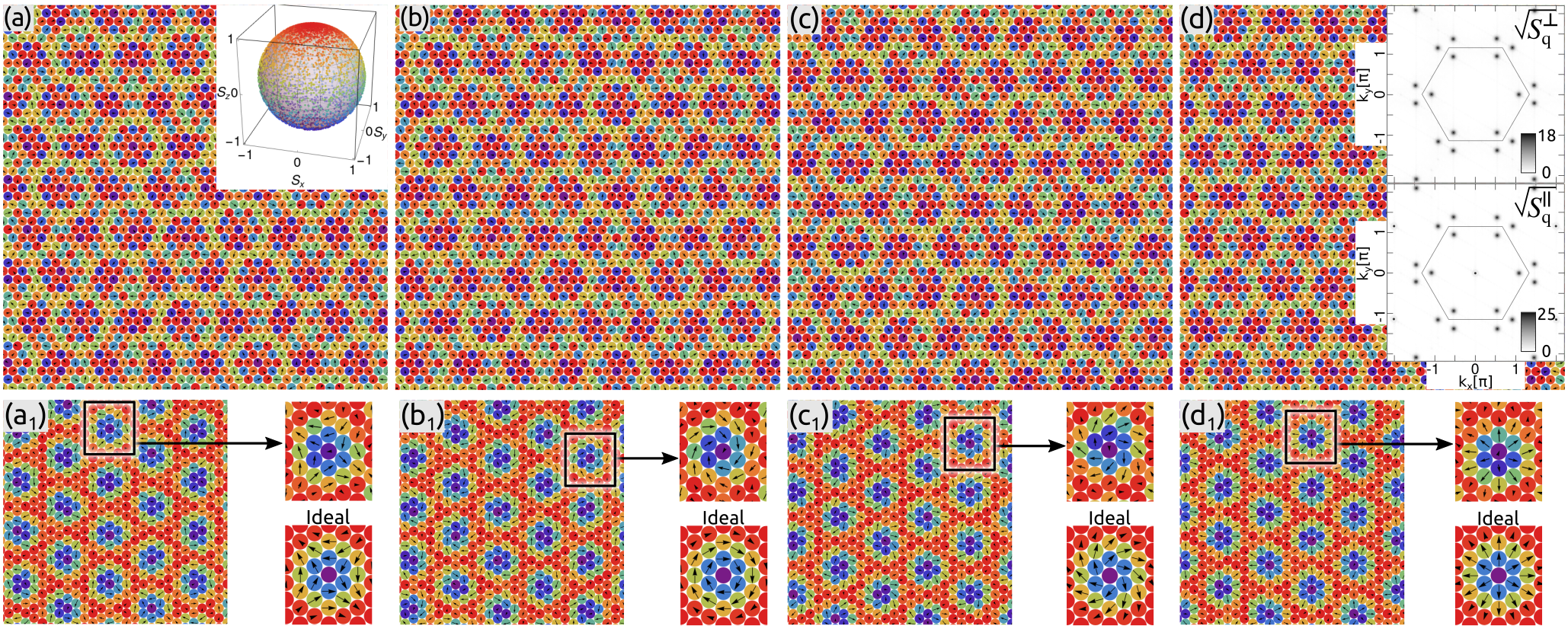}
\caption{\label{fig:AFsnaps} (Color online) Typical complete spin configurations for the antiferromagnetic antiskyrmion (panel  (a)) and skyrmions  (panels (b)-(d)) lattices stabilized for $J_2=0.3$, $J_3=0.16$, $T=3\times 10^{-3}$ and $B=3.52$. In  panel (a) we show the corresponding spherical snapshot while in the insets of panel (d) we show  the structure factors $\sqrt{S^{\perp}_{\qv}}$ and $\sqrt{S^{||}_{\qv}}$, presenting the characteristic triple-q structure. Panels (a$_1$)-(d$_1$) show one of the three $\sqrt{3} \times \sqrt{3}$ sublattices displaying the presence of a ferromagnetic antiskyrmion/skyrmion lattice. A comparison between an antiskyrmion/skyrmion obtained from simulations and an ``ideal'' one from an analytical  parametrisation is presented, to show the different topological charges $Q$ and helicities $\gamma$; from left to right ((a$_1$) to (d$_1$)): ``Bloch'' AF-ASkL ($Q=1,\gamma=-\pi/2$), Bloch AF-SkL ($Q=-1,\gamma=-\pi/2$), intermediate AF-SkL ($Q=-1,\gamma=-\pi/4$) and N\'eel AF-SkL ($Q=-1,\gamma=0$) lattices.}
\end{figure*}

We take as a representative point $J_2=0.3,J_3=0.16$. Notice the relative magnitude between couplings: compared with previous works with ferromagnetic nearest neighbor exchange interactions\cite{Okubo2012}, here the additional $J_2$, $J_3$ couplings are significantly smaller than $J_1$ and all the couplings are antiferromagnetic. The temperature vs magnetic field phase diagram is summarized in Fig.~\ref{fig:PD-asky}(a), where the boundaries of the regions were obtained by combining different variables, mainly the specific heat and the scalar chirality. The most remarkable feature is a finite region in temperature and magnetic field where a spontaneous antiferromagnetic  skyrmion/antiskyrmion lattice is stabilized. This region is surrounded by a single-q region, similar to the one found at zero magnetic field ({\bf C} phase), which turns into a double-q phase at higher magnetic fields ({\bf F} phase). 
In Fig.~\ref{fig:PD-asky}(b-e) we show representative snapshots and their corresponding longitudinal and transverse structure factors ($\sqrt{S^{||}_{\qv}}$ and $\sqrt{S^{\perp}_{\qv}}$) of the single-q ({\bf C}) and double-q ({\bf F}) phases, where the selected $\qv$ peaks in $S^{\perp}_{\qv}$ are a subset of the six minima found in the LTA analysis for this region in parameter space (compare with Fig.~\ref{fig:PD-LTA} (d)) and in $S^{||}_{\qv}$ a uniform component at ${\bf q}=0$ is induced by the applied field.

We now focus on the details of the emergent triple-q phase. There are three remarkable characteristics in this phase. First, as mentioned above, the real space structure is associated with either skyrmions or antiskyrmions: there is a spontaneous symmetry breaking which can be seen in the sign of the scalar chirality. Second, the textures are not simple (ferromagnetic) skyrmion or antiskyrmion lattices: the antiferromagnetic couplings induce antiferromagnetic skyrmion and antiskyrmion lattices, formed by three interpenetrated $\sqrt{3}\times\sqrt{3}$ triangular sublattices \cite{Rosales2015,Zukovic1,Zukovic2}. Third,  these antiferromagnetic skyrmions or antiskyrmions do not have a fixed helicity, they may be Bloch, N\'eel or intermediate (between Bloch and N\'eel type)\cite{BeyondSky,lin2015skyrmion,kim2018asymmetric}.  Other types of interactions, usually present in real materials, may also fix the helicity and vorticity, such as spin anisotropy \cite{Hayami2022-b} and dipolar interactions \cite{Utesov2022}. Typical snapshots of the full-lattice, one sublattice, and the corresponding structure factor are presented in Fig.~\ref{fig:AFsnaps}. In panels (a$_1$) to (d$_1$) we compare one single topological structure obtained from the simulations with an analytical parametrisation \cite{Osorio3} where we have changed the values of the topological charge $Q$ and the helicity $\gamma$ to show examples of the different types of textures stabilized for the same set of parameters.

\begin{figure}[h!]
\includegraphics[width=0.5\textwidth]{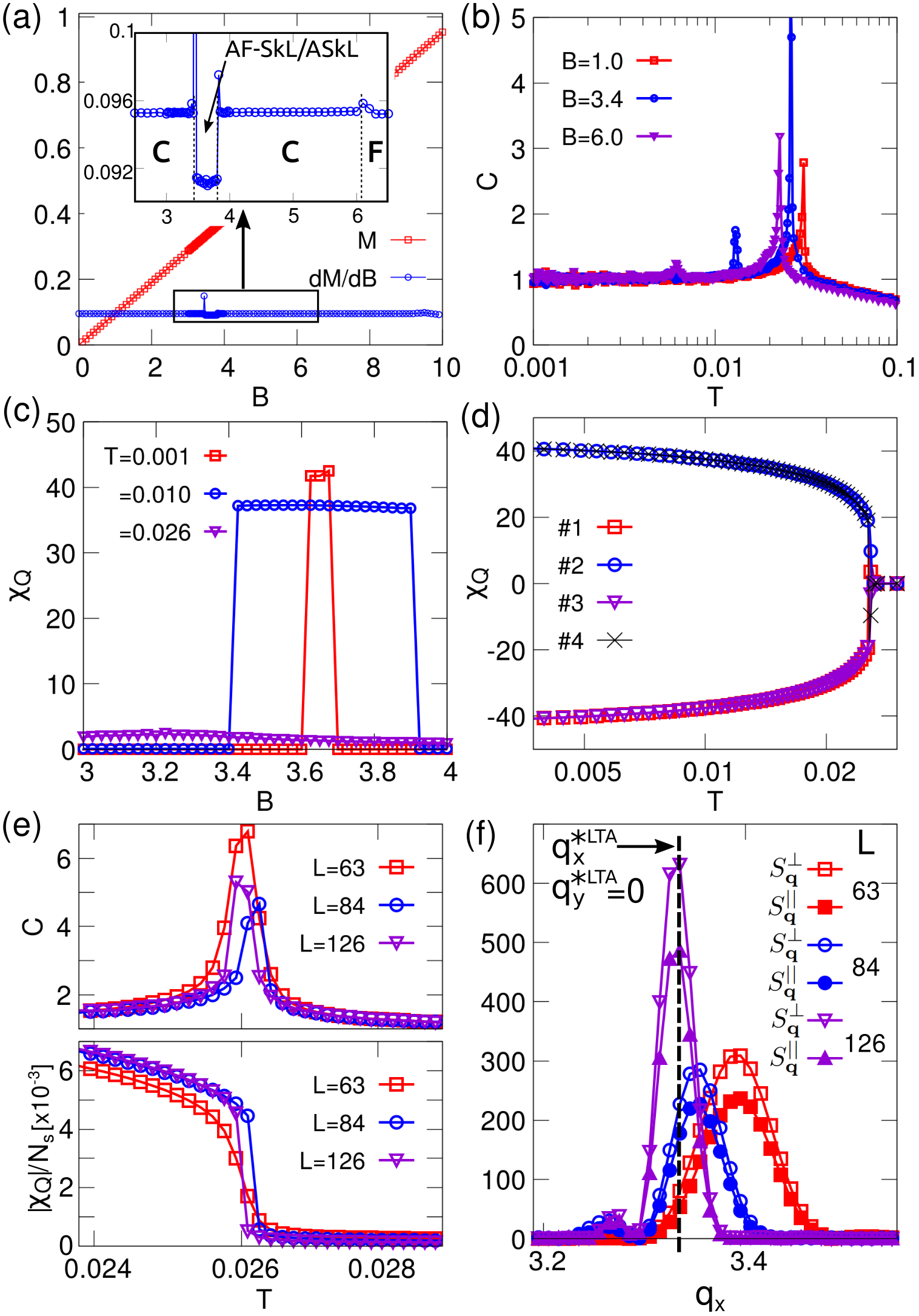}
\caption{\label{fig:AFvariables}(Color online) Thermodynamic variables calculated with MC simulations for $J_2=0.3,J_3=0.16$ . (a) Magnetization $M$ (red) and its derivative (blue) as a function of the magnetic field at $T=5\times 10^{-3}$. The inset zooms on the $dM/dB$ curve, showing a dip that indicates the AF-SkL/ASkL region (b) Specific heat as a function of temperature for three values of the magnetic field. The AF-SkL/ASkL phase is found at $B=3.4$, in the region between the two  peaks (c) Mean value of the absolute value of the chirality as a function of the magnetic field for three different temperatures (d) Chirality as a function of temperature at $B=3.5$ for four different MC realizations and $L=63$. (e) Specific heat and chirality density as a function of temperature for three different system sizes ($L=63,84,126$) for $B=3.5$. (f)  Dependence of the peaks of the transverse ($S^{\perp}_{\qv}$) and longitudinal  ($S^{||}_{\qv}$) structure factors with system size in the AF-SkL/ASkL region ($B=3.5, T=4\times 10^{-3}$). The dashed vertical lines indicates the LTA analytical solution ${\bf q}^{*LTA}=(3.335,0)$.}
\end{figure}

This phase has clear signatures in different thermodynamic quantities. In Fig.~\ref{fig:AFvariables} (a) we present the magnetization curve at $T=10^{-3}$, where the inset shows the changes in the slope of the curve indicating that the system enters  different phases. Specific heat curves as a function of temperature at three different magnetic fields $B=1.0,3.4,6.0$ are presented in (b). The AF-SkL/ASkL  phase at $B=3.4$ is defined as the region between the two peaks. Most importantly, in Fig.~\ref{fig:AFvariables} (c) and (d) we present the behavior of the topological parameter, the scalar chirality: mean value of its absolute value as a function of magnetic field for three different temperatures in (c), and the chirality for four different MC realizations as a function of $T$ for $B=3.5$ in (d), where it can be seen that indeed it takes either  negative or positive values, corresponding to either skyrmions or antiskyrmions lattices. To confirm the stability of the AF-SkL/ASkL  phase with system size, in panel (e) we plot the specific heat and the absolute value of the chirality per site for three different system sizes at $B=3.5$. In panel (f) we show, for $B=3.5$ and $T=4\times10^{-3}$, the dependence with  system size of $S_q^{\perp}$ and $S_q^{\parallel}$  measured along the line going through the peak position, and compare it with the LTA solution  ${\bf q}^{*LTA}=(3.335,0)$ (indicated with a vertical black dashed line). It can be seen that, as the system size is increased, the peaks sharpen and are closer to the LTA result. The behaviour with system size is also consistent with  quasi-long range order, as discussed in Ref. [\onlinecite{Shimokawa2019}]. 

As in the previous subsection, as possible material realization hosting AF-SkL/ASkL phase we can mention the family of materials Cr/MoS$_2$, Fe/MoS$_2$, and Fe/WSe$_2$ with triangular geometry,  where recently was predicted that skyrmion lattices could appear even for relatively weak DMI\cite{Fang2021}. 

\section{SUMMARY and CONCLUSIONS}

We have studied a pure antiferromagnetic isotropic model in the triangular lattice where the combination of frustrating interactions,  external magnetic field, and temperature induces a variety of multiple-q and spiral spin liquid phases. We approach this study through two complementary techniques. First, we explore the possible ground states using the Luttinger-Tisza approximation, which is a strong analytical tool to identify regions with possible exotic phases. Then, we resort to large-scale Monte-Carlo simulations, combining Parallel Tempering and the Metropolis algorithm with overrelaxation, to study the effect of temperature.

At zero temperature and zero magnetic field, our LTA analysis shows seven distinct phases in the $J_2-J_3$ space, which we classify according to the position of the ordering wave vectors $\qv$ at the energy minima in the Brillouin zone. There are two regions where our results match previous theoretical studies \cite{Lhuillier2011} and the energy minima lie in the $K$ and $M$ points of the Brillouin Zone. Then, further exploring parameter space, we find two broad regions  with possible triple-q topological phases, with six incommensurate $\qv$ minima. The difference between these phases lies in the position of these minima: in one case they lie in the line between the $\Gamma$ and the $K$ points, in the other between $\Gamma$ and $M$ points. There is a fifth small region where there are 12 energy minima in the borders of the BZ, between the $K$ and $M$ points. Moreover, there is a particular line $J_2=2J_3>1/4$ where we find two types of states with semi-extensive degeneracy. For $1/4<J_2=2J_3<1/3$, the minima are closed lines that encircle the $K$ points in the Brillouin Zone. For $J_2=2J_3>1/3$, the minima reside in a ring centered at the $\Gamma$ point. Therefore, this LTA study suggests two types of possible exotic behavior with temperature: the incommensurate triple-q phases may give rise, with the addition of an external field, to skyrmion-like non-trivial topological textures, and the degenerate lines indicate possible spiral spin liquids.

In order to study the emergent phenomena with thermal fluctuations, we resort to large-scale Monte-Carlo simulations. At zero magnetic field, we first find that in the region where the minima from LTA lay in the $M$ points, there is an order-by-disorder state selection to single-q states with antiferromagnetic (and thus collinear) stripe ordering. Then, in the incommensurate regions, we find  single-q phases, where the selected $\qv$ are one of the triple-q minima found with the LTA. An interesting behavior arises in the $J_2=2J_3>1/4$ line, where thermal fluctuations drive the system from two types of spiral spin-liquids to  single-q orderings.

Most importantly, we find that temperature and magnetic field stabilize a spontaneous topological phase, where either antiferromagnetic skyrmion or antiskyrmion lattices, and non-fixed helicity  (Bloch, N\'eel or intermediate) are found.  We show the signatures of this phase in observables such as  the magnetization and the specific heat, and use the scalar chirality as the parameter to illustrate the formation of either skyrmion or antiskyrmions. These configurations are seen in reciprocal space as triple-q phases with six incommensurate peaks lying between the $\Gamma$ and $K$ points. 

In conclusion, we see that the competition between isotropic antiferromagnetic interactions in a frustrated lattice, with dominant nearest neighbor exchange couplings, is also a mechanism to stabilize antiferromagnetic skyrmion-like lattices under external magnetic fields, without antisymmetric or  anisotropic additional interactions. Since the model retains rotation symmetries in the plane perpendicular to the field, there are different types of topological phases, combining two types of crystals of quasiparticles (skyrmions or antiskyrmions), which is reflected in the chirality, and different types of helicities, Bloch, N\'eel or intermediate. Beyond this model, additional perturbative interactions which are usually present in real materials, such as single-ion, bond anisotropy,  and dipolar interactions \cite{Utesov2022}, may favour antiferromagnetic topological structures with fixed helicity and topological charge. The effect of colective excitations, such as phasons \cite{Tatara2014,Wang2022}, may also play a role in the stabilization of these textures. We trust that this work further contributes to the exploration of non-trivial topological phases, and  their realization in frustrated materials.

\section*{Acknowledgments} 

M. M. and M. \v{Z}.  are supported by the grants of the Slovak Research and Development Agency (Grant No. APVV-20-0150) and the Scientific Grant Agency of Ministry of Education of Slovak Republic (Grant No. 1/0531/19). F. A. G. A. and H. D. R. are partially supported by CONICET (PIP 2021-112200200101480CO),  SECyT UNLP PI+D  X893 and PICT-2020-SERIEA-03205. F. A. G. A. acknowledges support from PICT 2018-02968.

\bibliographystyle{apsrev4-1}

%

\clearpage
\onecolumngrid

\end{document}